\documentclass[twocolumn,prb]{revtex4}
\usepackage{graphicx}
\usepackage{bm}
\usepackage{amssymb}
\usepackage[centerlast]{subfigure}

\begin{document}

\title{Fractal clustering of inertial particles in random flows}

\author{J\'{e}r\'{e}mie Bec}

\email{bec@obs-nice.fr}

\affiliation{Institute for Advanced Study, Einstein Drive, Princeton,
  New Jersey 08540.\\ Lab.{\ }G.-D.{\ }Cassini, Observatoire de la
  C{\^{o}}te d'Azur, BP 4229, 06304 Nice Cedex 4, France.}

\begin{abstract}
  \noindent It is shown that preferential concentrations of inertial
  (finite-size) particle suspensions in turbulent flows follow from
  the dissipative nature of their dynamics. In phase space, particle
  trajectories converge toward a dynamical fractal attractor. Below a
  critical Stokes number (non-dimensional viscous friction time), the
  projection on position space is a dynamical fractal cluster; above
  this number, particles are space filling. Numerical simulations and
  semi-heuristic theory illustrating such effects are presented for a
  simple model of inertial particle dynamics.
\end{abstract}
\maketitle

\noindent Dust particles, droplets, bubbles and various impurities
advected by turbulent flow have usually a finite size and a mass
density differing from that of the carrier fluid. Contrary to passive
tracers, whose dynamics is conservative (when the flow is
incompressible), the dynamics of such \emph{inertial particles}\/ is
rendered dissipative by the Stokes drag, which can lead to strongly
inhomogeneous spatial distributions. The full statistical description
of such preferential concentrations is still an open question with
many natural and industrial applications, such as the growth of rain
drops in subtropical clouds,\cite{ffs02} the formation of
planetesimals in the early Solar system,\cite{bcps99} optimization of
combustion processes and the coexistence problems between several
species of plankton.\cite{kpstt00}

Maxey and Riley\cite{mr83} derived an equation for the motion of a
rigid spherical particle embedded in an incompressible flow. They
assume that (i) the particle is smaller than the smallest turbulent
scale of the carrier flow and that (ii) the Reynolds number associated
to its size and its relative velocity with respect to the fluid is
sufficiently small to approximate the surrounding flow by a Stokes
flow. Then, the forces exerted on the particle are buoyancy, the force
due to the undisturbed flow, the Stokes viscous drag, the added mass
effect and the Basset--Boussinesq history force. Because of the
complexity of the resulting equation of motion, simpler models are
generally used. For instance, when the Stokes drag is very strong, the
dynamics is close to that of passive tracer particles and the
discrepancy can be captured by a spatial Taylor expansion, leading to
a model in which the particles are advected by a synthetic flow
comprising a small compressible component.\cite{ekr96,bff01} What singles
out the model proposed here is its ability to take into account the
full phase-space dynamics of the particles and to capture the
essential features of their dissipative motion. We are interested in
the ``Batchelor r\'{e}gime'' of the particles, meaning that we focus
on spatial scales smaller than the Kolmogorov dissipation scale
$\eta$. After rescaling of space, time and velocity respectively by
factors $\eta$, $\eta^2/\nu$ and $\nu/\eta$, and assuming that
the particle velocity is sufficiently close to that of the fluid, the
Newton equation satisfied by its trajectory ${\bf X}(t)$ reduces to
\begin{equation}
  \ddot{\bf X} = \beta \frac{d}{dt}\left[\bm u({\bf X}, t)\right] -
    \frac{1}{{\rm S}_\eta} \left[\dot{\bf X}-\bm u({\bf X}, t)\right],
  \label{eq:newton}
\end{equation}
where $\beta \equiv 3\rho_f/(\rho_f+2\rho_p)$ is the added-mass factor
($\rho_f$ and $\rho_p$ are the fluid and the particle mass densities)
and ${\rm S}_\eta \equiv a^2/(3\beta\eta^2)$ is the \emph{Stokes
number} associated to the dissipative scale of the carrier flow ($a$
being the particle radius). Introducing the \emph{co-velocity} ${\bf
V} \equiv \dot{\bf X} - \beta\bm u ({\bf X},t)$, the equation of
motion can be interpreted in terms of the $(2\times d)$-dimensional
dynamical system
\begin{eqnarray}
  \dot{\bf X} &=& \beta \bm u ({\bf X},t) + {\bf V},
  \label{eq:newton1} \\
  \dot{\bf V} &=& \frac{1}{{\rm S}_\eta} \left[ (1-\beta) \bm u({\bf
  X},t) - {\bf V} \right]. \label{eq:newton2}
\end{eqnarray}
The motion of the particles is clearly dissipative, even if the
carrier flow is itself incompressible\,: indeed, when $\nabla\cdot \bm
u =0$, the contraction rate in phase space reduces to $- d/{\rm
S}_\eta$ and is strictly negative, inducing a uniform contraction. As
a consequence, the long-time dynamics of the particles is
characterized by the presence of an attractor, that is a dynamical
fractal set of the phase space toward which the trajectories $({\bf
X}(t), {\bf V}(t))$ converge. Important information on the dynamical
system (\ref{eq:newton1})-(\ref{eq:newton2}), regarding stability,
Lyapunov exponents, etc is obtained from the linearized equation
governing the separation ${\bf R}(t) \equiv (\delta{\bf X}(t),
\delta{\bf V}(t))$ between two infinitesimally close trajectories of
the phase space. For scales within the viscous scale of turbulence,
the velocity field $\bm u$ can be considered spatially smooth and the
separation ${\bf R}(t)$ obeys the linear differential equation
\begin{equation}
\dot{\bf R} = {\cal M}_t \,{\bf R}, \qquad {\cal M}_t\equiv \left[
\begin{array}{lr} \beta \bm \sigma(t) & {\cal I}_d \\
\displaystyle\frac{1-\beta}{{\rm S}_\eta} \bm \sigma(t) &
\displaystyle \frac{1}{{\rm S}_\eta} {\cal I}_d \end{array} \right],
\label{eq:tangentsys}
\end{equation}
where $\bm \sigma$ is the strain matrix of the carrier flow along the
path of a reference particle\,: $\sigma_{ij}(t) \equiv \partial_j u_i
({\bm X}(t), t)$, and ${\cal I}_d$ is the $d$-dimensional identity
matrix. A full stability analysis of the dynamics can easily be
done;\cite{b02} it relates the eigenvalues of the evolution matrix
${\cal M}_t$ to that of the stress tensor $\bm\sigma$ representative
of the local structure of the carrier flow. In both two and three
dimensions, this leads to distinguishing between heavy ($\beta<1$) and
light particles ($\beta>1$): the former are usually ejected from the
elliptic regions, while the latter may cluster there in a pointwise
manner. We therefore expect the vortex cores to be regions of high
concentration of light particles and of low concentration of heavy
particles, feature which is generally observed in experiments and
simulations (for a review, see Eaton and Fessler\cite{ef94}).

We focus here on suspensions of particles with a volume fraction
sufficiently small to neglect their interactions and
collisions. Typically, the phase-space attractor on which the
particles concentrate is a fractal object which may be characterized
by various dimensions, in particular a non-random Hausdorff dimension
$d_H$. As the position of the particles is obtained by projection from
the $2d$-dimensional phase space onto the $d$-dimensional physical
space, the convergence to the attractor is responsible for strong
inhomogeneities in the large-times distribution of particles. More
precisely, a standard result of the geometrical theory of fractal
sets\cite{f86} states that if $d_H<d$, the distribution of particles
in the physical space is itself a fractal set with Hausdorff dimension
$d_H$, whereas if $d_H>d$, the particles fill the whole space. Hence,
depending on the value of the dimension $d_H$ of the attractor, two
different r\'{e}gimes are distinguished.  Clearly, the dimension of
the attractor is a function of the Stokes number ${\rm S}_\eta$ and of
the added mass parameter $\beta$, and generally also depends on the
statistical properties of the velocity of the carrier fluid. Leaving
aside this latter dependence, let us first note what can be easily
inferred on the behavior of $d_H$ as a function of ${\rm S}_\eta$ and
$\beta$. First, in the limit of vanishing Stokes numbers, there is a
reduction of dimensionality and the dynamics of simple tracers is
recovered. An initially uniform distribution of particles remains
uniform and we hence have $d_H\to d$. Next, for very large Stokes
numbers, the particles are less and less influenced by the carrier
fluid and their motion becomes ballistic. They thus fill the whole
phase space and we have $d_H\to 2d$. In between these two asymptotic
r\'{e}gimes, although it is not obvious that the dimension $d_H$ can
fall below $d$, we shall actually show that there is a whole range of
Stokes numbers for which $d_H<d$ and, thus, preferential concentration
on fractal clusters occurs in physical space.

Finding theoretically or numerically the Hausdorff dimension $d_H$
of the attractor is not, in general, a simple task: its determination
demands a full understanding of the global dynamics and its numerical
measurement requires very large numbers of particles. To obtain a
simple estimate of the attractor dimension, Kaplan and
Yorke\cite{ky79} proposed to use the \emph{Lyapunov dimension}. It is
given by the Lyapunov exponents $\lambda_1>\dots>\lambda_{2d}$ which
measure the exponential growth of infinitesimal distances, surfaces
and volumes in the phase space and are expressed in terms of the
limiting singular values of the Jacobi matrix, i.e.
\begin{equation}
  \lambda_j \equiv \lim_{t\to\infty} \frac{1}{t} \ln \left|{\cal J}_t
  \bm e_j\right|,\quad {\cal J}_t \equiv {\cal T}\!\!\exp \int_0^t
  {\cal M}_s ds,
  \label{eq:deflyapandjacobi}
\end{equation}
where the $\bm e_j$'s are the eigenvectors of the symmetric matrix
${\cal J}_t^{\rm T}{\cal J}_t$ and ${\cal T}\!\!\exp$ denotes the
time-ordered exponential. The Lyapunov dimension is defined as
\begin{equation}
  d_L \equiv j - \frac{\lambda_1+\dots+\lambda_j}{\lambda_{j+1}}, 
  \label{eq:deflyapdim}
\end{equation}
where $j$ satisfies $\lambda_1+\dots+\lambda_j \ge 0$
and $\lambda_1+\dots+\lambda_{j+1} < 0$. Beside being a simple
estimate of the dimension of the attractor, it was actually
shown\cite{do80} that the Lyapunov dimension gives a rigorous upper
bound for the Hausdorff dimension $d_H$ of the attractor.

\begin{figure}[htbp]
\centerline{\includegraphics[width=0.4\textwidth]{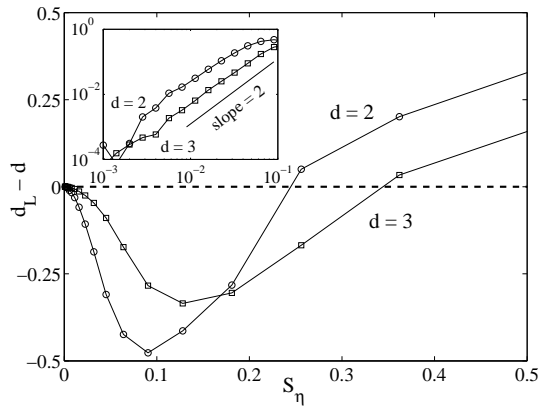}}
\caption{\footnotesize \label{fig:dimension} Difference between the
  Lyapunov dimension $d_L$ of heavy particles ($\beta=0$) and the
  physical space dimension $d$, versus the Stokes number ${\rm
  S}_\eta$ (circle: $d=2$, square: $d=3$). The critical Stokes number
  corresponds to the value for which $d_L=d$. Upper-left inset\,: same
  in log-log coordinates showing a quadratic behavior at small Stokes
  numbers.}
\end{figure}
\vspace{-15pt}
\begin{figure}[htbp]
  \centerline{\subfigure[\label{fig:cluster} ${\rm S}_\eta = 10^{-2}$]
    {\includegraphics[width=0.23\textwidth]{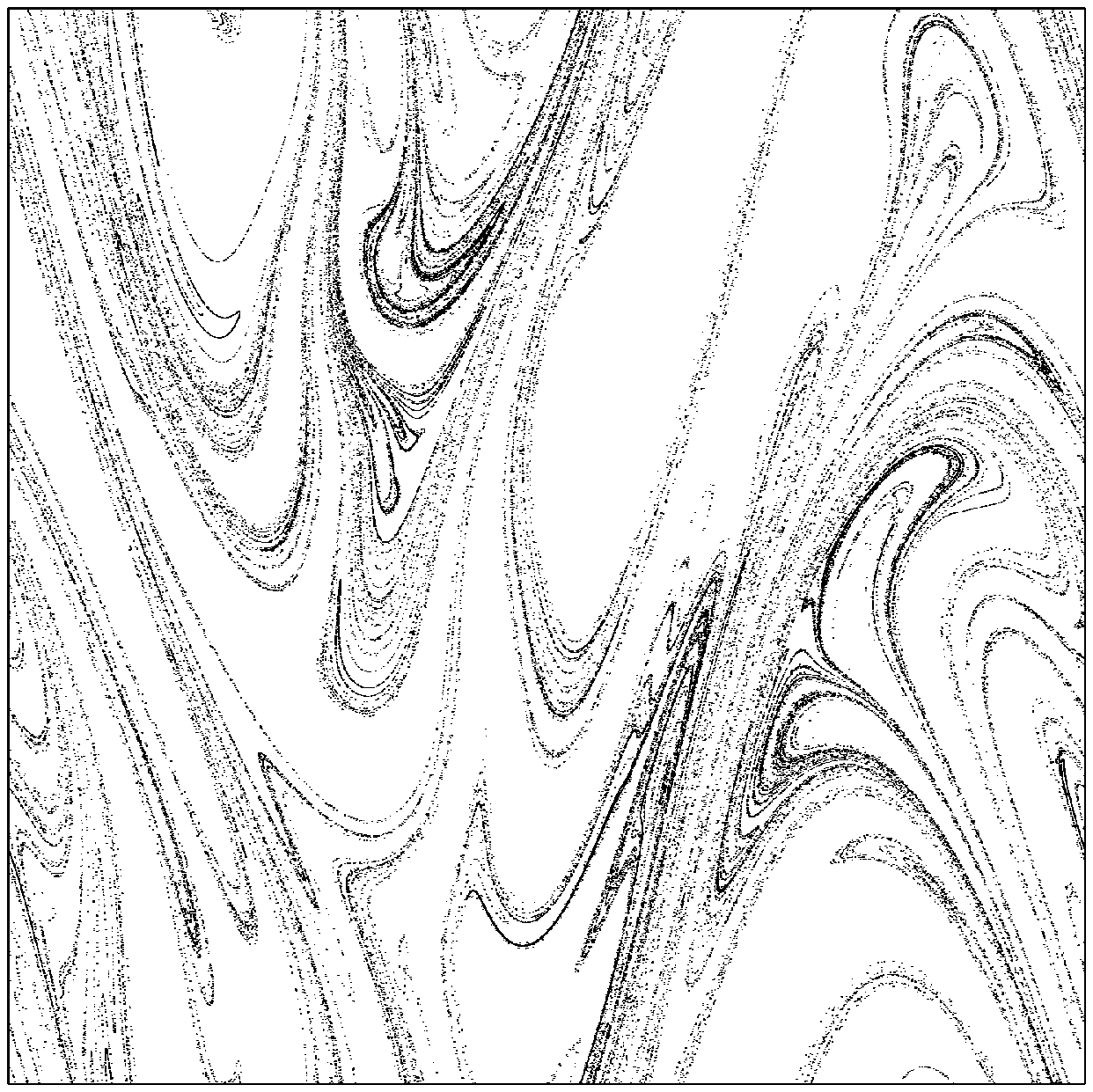}} \quad
    \subfigure[\label{fig:notcluster} ${\rm S}_\eta = 1$]
              {\includegraphics[width=0.23\textwidth]{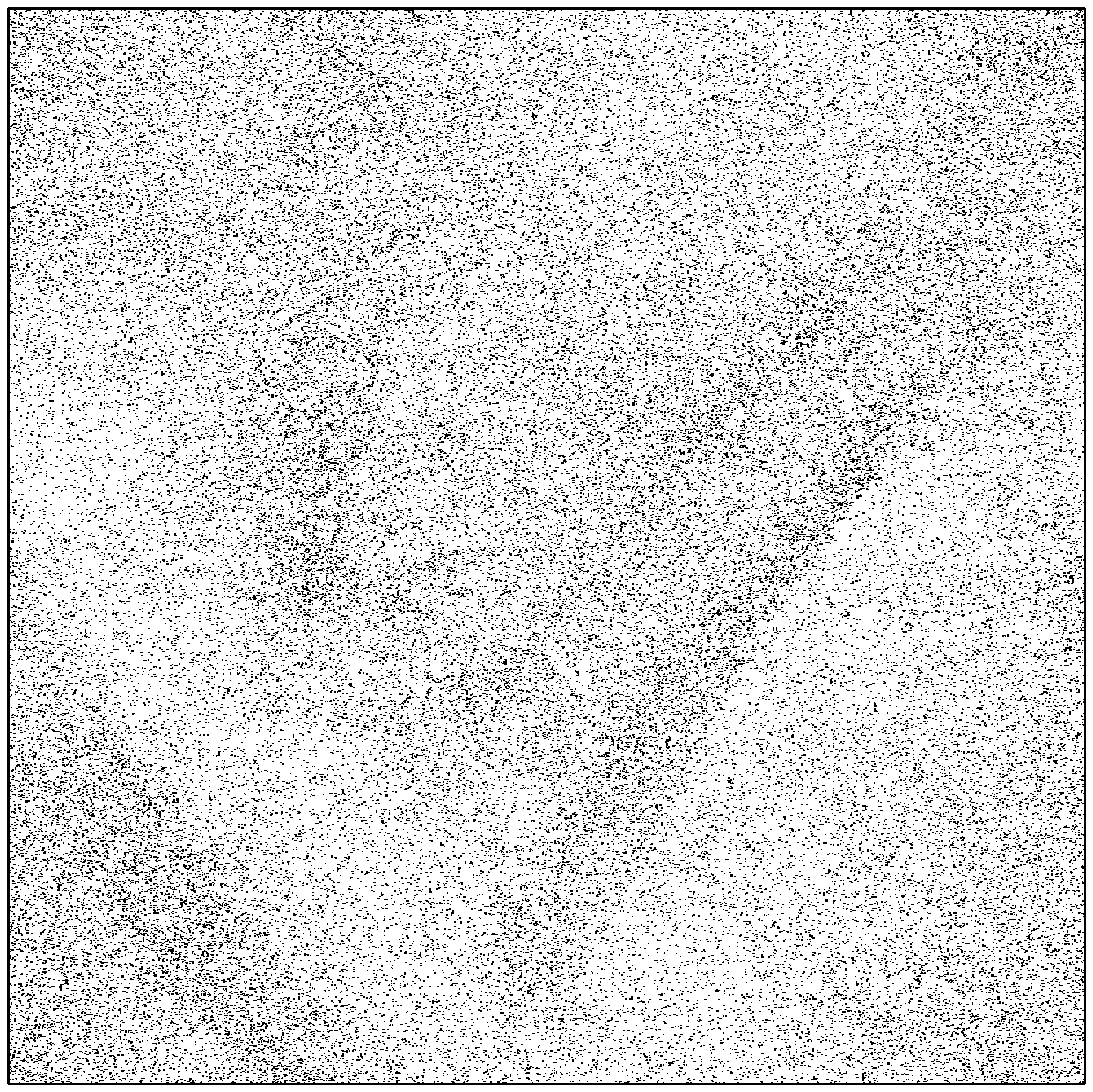}}}
  \vspace{-10pt}
  \caption{\label{fig:2regimes}\footnotesize Snapshots of the position
    of $N=10^5$ heavy particles ($\beta=0$) associated to two
    different Stokes numbers: (a) smaller than the critical value, for
    which the particles form fractal clusters, and (b) larger than the
    critical value for which they fill the whole domain. The carrier
    incompressible flow is generated randomly by 4 independent modes
    with a finite correlation time.}
\end{figure}
We performed numerical experiments in two and three dimensions for a
space-periodic carrier flow generated randomly by the superposition of
few independent Gaussian Fourier modes with a correlation time of the
order of unity (this specific form for the carrier flow was considered
by Sigurgeirsson and Stuart\cite{ss02a} who proved the existence of a
random dynamical attractor). The Lyapunov exponents are calculated by
the use of the standard technique of Benettin {\it et
al.}\cite{bggs80}, and the resulting Lyapunov dimension is represented
both for $d=2$ and $d=3$ in Fig.~\ref{fig:dimension} for the case of
very heavy particles ($\beta=0$).  Two important observations can be
made from these simulations. First, fractal clustering of particles
already occurs at very small Stokes number where the Lyapunov
dimension behaves as $d_L\simeq d - C {\rm S}_\eta^2$ with $C>0$. This
quadratic behavior near zero was predicted for the correlation
dimension by Balkovsky {\it et al.}\cite{bff01} using the method of
the synthetic compressible flow cited earlier, as an approximation at
low Stokes numbers. The second observation is the presence of a
critical value for the Stokes number below which the attractor
dimension is smaller than $d$ and where the particles form fractal
clusters in the physical space.  The two r\'{e}gimes corresponding to
different values of the Stokes number are illustrated for $d=2$ in
Fig.\ \ref{fig:2regimes}. When the Stokes number is below the critical
value (a), the particles concentrate onto a fractal set and both very
dense and almost empty regions appear. On the contrary, for a Stokes
number above the threshold (b), the particles fill the whole domain,
albeit with a non uniform density.

\begin{figure}[htbp]
\centerline{\includegraphics[width=0.37\textwidth]{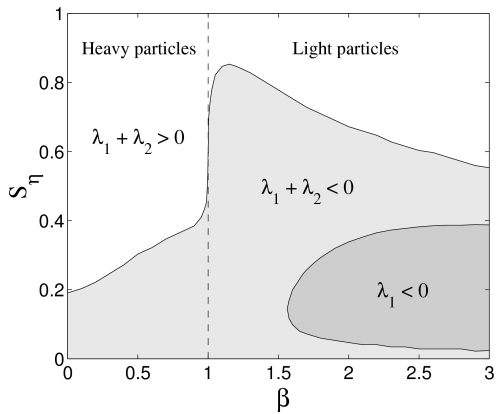}}
\caption{\footnotesize \label{fig:diagramme} Phase diagram in the
  parameter space $(\beta, {\rm S}_\eta)$ for the two-dimensional
  case, representing the three different r\'{e}gimes classified by the
  behavior of the Lyapunov exponents.}
\end{figure}
\vspace{-10pt} When the particles have a finite mass ($\beta\neq0$),
there also exists a critical Stokes number for their concentration
onto fractal clusters. Figure \ref{fig:diagramme} shows for $d=2$ the
phase diagram obtained numerically which divides the parameter space
$(\beta, {\rm S}_\eta)$ between three different r\'{e}gimes.  When the
sum of the two largest Lyapunov exponents is negative (light-gray
domain), we have $d_L<d$ and the particles form fractal clusters in
the physical space. When the sum is positive (white domain), we have
$d_L>d$ and the particles fill the whole domain. The third case occurs
only for particles lighter than the fluid and corresponds to a
negative largest Lyapunov exponent (dark gray area). The particles
form pointwise clusters and, when the domain is bounded, they all
converge to a single trajectory.

We now present a heuristic argument, which we already gave in the case
$\beta=0$ of heavy particles,\cite{bd02} and which predicts the
threshold in Stokes number. It relies on the use of the
\emph{stability exponents}, that are the exponential growth rates of
the eigenvalues of the Jacobi matrix ${\cal J}_t$ defined in
(\ref{eq:deflyapandjacobi}). Using Browne's theorem which bounds the
singular values of a square matrix by its eigenvalues, a necessary
condition for the fractal clustering of particles is that the sum of
the $d$ largest stability exponents is positive. For heavy particles
($\beta<1$), this sum can be estimated from the local analysis of the
dynamics. First, since such particles tend to cluster within the
hyperbolic regions of the flow, they are spending there a fraction of
time much larger than in the elliptic regions. Let us assume that the
relationship between the eigenvalues of the evolution matrix ${\cal
M}_t$ and those of the stress tensor $\bm\sigma(t)$ can be extended to
the stability exponents, at least as an approximation; we can then
derive a necessary condition for the presence of fractal clusters in
the physical space. For $d=2$, this condition can be easily written as
\begin{equation}
  {\rm S}_\eta \le \frac{1}{\lambda_f\beta^2} \left( \beta-2
  +2\sqrt{1-\beta+\beta^2} \right ),
  \label{eq:condcluster2d}
\end{equation}
where $\lambda_f$ is the (non-dimensional) largest Lyapunov
exponent associated to the carrier velocity field and calculated along
the trajectory of a simple fluid particle. It was easily verified that
this bound is compatible with what is observed numerically in
Fig.~\ref{fig:diagramme}.

It is often stated in the literature that the clustering of inertial
particles is essentially due to the presence of coherent structures in
turbulence (see., e.g., Squires and Eaton\cite{se91}). It is indeed
generally assumed that the structures with long life times appearing
in the flow are responsible for a deterministic motion of the particle
leading to their concentration inside or outside the
vortices. Although this argument, based on the local structure of the
carrier flow, allowed us to find an upper bound for the critical value
of the Stokes number, it is important to stress that preferential
concentrations of particles arise solely as a consequence of the
dissipative character of the motion. Indeed, we have also performed
simulations with carrier flows which are delta-correlated in time and
that are thus completely devoid of structure.\cite{b02} These
simulations show that the dependence of the Lyapunov dimension on the
Stokes number ${\rm S}_\eta$ is very similar to that obtained in
Fig.~\ref{fig:dimension}. The main difference is the behavior of $d_L$
as ${\rm S}_\eta\to0$: in the delta-correlated case, the Lyapunov
dimension tends linearly, and not quadratically, to the space
dimension $d$. In the delta-correlated case, preferential
concentrations are actually stronger than for a finite correlation
time, contradicting the mechanism frequently invoked to explain
concentrations.

Of course, it is not sufficient to know that the particles are
concentrated in fractal objects; a fuller statistical description of
their distribution is desirable. In particular, for a quantitative
description of their spatial intermittency, one needs their
multifractal properties (scaling exponents of the various moments of
the mass contained in a sphere of radius $r$; see e.g.\ Chap.\ 8 of
the book by Frisch\cite{f95}). Preliminary results, for heavy
particles, indicate that strong spatial intermittency can be present,
even when the Lyapunov dimension is just slightly below that of the
physical space. Let us also mention that more quantitative results are
likely to be within reach using multi-time methods in the asymptotics
$S_\eta\ll1$. Small Stokes numbers are of considerable interest since
in most natural and industrial situations, this is the case. A last
remark concerns the extension of this approach to the case of real
turbulent flows. To confirm the existence of a threshold for fractal
clustering of inertial particles, one needs to resolve scales which
are much smaller than the Kolmogorov dissipation scale $\eta$. This
is quite a challenge, both for laboratory and numerical experiments.

\medskip
I am deeply grateful to K.\ Domelevo for initiating my work on
inertial particles and to U.\ Frisch for his continuous
support. During this work, I have benefited from stimulating
discussions with A.\ Celani, M.\ Cencini, P.~Horvai, K.\ Khanin, M.\
Stepanov and P.\ Tanga. Part of this research was possible thanks to
the support of the \emph{Soci\'{e}t\'{e} de Secours des Amis des
Sciences}\/ which is warmly acknowledged. Part of this work was also
supported by the National Science Foundation under agreement No.\
DMS-9729992 and by the European Union under contract
HPRN-CT-2000-00162. Numerical simulations were performed in the
framework of the SIVAM project at the Observatoire de la C{\^{o}}te
d'Azur.

\end{document}